\documentclass[twoside]{dis07}
\usepackage[latin1]{inputenc}
\usepackage[dvips]{graphicx,epsfig,color}
\usepackage{wrapfig,rotating}
\usepackage{amssymb,amsmath,array}

\begin{document}
\title{Photoproduction of Hadron Pairs at Fixed-Target Experiments}

\author{Christof Hendlmeier$^1$, Marco Stratmann$^2$ and Andreas Schäfer$^1$
%
\vspace{.3cm}\\
%
1- Universität Regensburg - Institut für Theoretische Physik, D-93040 Regensburg - Germany
%
\vspace{.1cm}\\
2- Radiation Laboratory - RIKEN, Wako, Saitama 351-0198 - Japan
}

\maketitle

\begin{abstract}
We consider the photoproduction of two hadrons in polarized lepton-nucleon
collisions in the framework of perturbative QCD at the next-to-leading order
accuracy \cite{url}. After illustrating how to
obtain the experimentally relevant observables, a phenomenological
study of the photoproduction of hadron pairs at high transverse momenta is
presented. We show theoretical predictions for the relevant cross sections 
at COMPASS and HERMES kinematics as well as
theoretical uncertainties.
\end{abstract}

\section{Motivation}

After more than 25 years of studying polarized deep-inelastic lepton-nucleon 
scattering (DIS) the prime question is still how the proton spin-$\frac{1}{2}$ is 
composed of the spins and orbital angular momenta of its constituents, quarks 
and gluons. The single most important result is the finding that quarks spins 
contribute only little - about $25\%$ - to the proton spin \cite{rith}. 
The measurement 
of $\Delta g(x,\mu)$, the polarized gluon distribution of the proton, is the 
next logical step to clarify the spin puzzle, since it turns out that in the light 
cone gauge the first moment of $\Delta g(x,\mu)$ has the interpretation of the total 
contribution of the gluons spin to the proton's spin-$\frac{1}{2}$.
The extraction of $\Delta g$ in polarized DIS is, however, very difficult as it 
contributes 
only via scaling violations and at higher orders in the strong coupling constant $\alpha_s$. 
Therefore 
the prime goal of all current experiments with polarized beams is to determine 
$\Delta g$ directly.

In particular at the Relativistic Heavy Ion Collider (RHIC) at Brookhaven National 
Laboratory (BNL), many different processes can be studied where $\Delta g$ enters 
dominantly already at the lowest order (LO) approximation of perturbative QCD (pQCD): 
for example prompt photon and heavy flavor production, jet and single-inclusive hadron 
production as well as di-hadron production \cite{rhic1}. 
First results from the PHENIX and STAR collaboration at 
RHIC indicate 
that a large and positive gluon distribution is strongly disfavored in the probed 
region of momentum fractions $x$, $0.03\lesssim x\lesssim 0.2$ \cite{rhic2}.

In addition to RHIC, further information on the spin structure of nucleons can be 
obtained by fixed-target experiments like COMPASS \cite{compass1} at CERN or HERMES 
\cite{hermes1} at DESY. One 
promising process for the determination of $\Delta g$ at the low energies 
available at fixed-target experiments turned out to be the production of hadron 
pairs at high transverse momenta $p_T$ \cite{bravar}. 
Some experimental results are already available \cite{results}.

\section{Technical framework}

We consider the spin-dependent inclusive photoproduction process
\begin{equation}
\vec{l}(p_l)+\vec{N}(p_N)\to l^\prime(p_{l^\prime}) H_c(p_c)H_d(p_d) X,
\end{equation}
where a longitudinally polarized lepton $\vec{l}$ with four momentum $p_l$ scatters 
off a longitudinally polarized nucleon $\vec{N}$ with four-momentum $p_N$ producing
two hadrons $H_c$ and $H_d$ with four momenta $p_c$ and $p_d$, respectively.
The two produced hadrons are assumed to have high transverse momenta $p_{T,c/d}$.
Making use of the factorization theorem, the polarized cross section can be written 
as a convolution of non-perturbative parton distribution and fragmentation functions
and hard, short-distance partonic cross sections:
\begin{eqnarray}
d\Delta\sigma&\equiv&\frac{1}{2}[d\sigma_{++}-d\sigma_{+-}]=\sum_{abcd}\int dx_a dx_b dz_c dz_d \Delta f^l(x_a,\mu_f)\Delta f^N(x_b,\mu_f)×\nonumber\\
&&d\Delta\hat{\sigma}^{ab\to cdX^\prime}(S,x_a,x_b,p_c/z_c,p_d/z_d,\mu_f,\mu_f^\prime,\mu_r)
D_c^{H_c}(z_c,\mu_f^\prime)D_d^{H_d}(z_d,\mu_f^\prime).
\label{xsec}
\end{eqnarray}
The sum runs over all possible partonic channels $ab\to cd$ with 
$d\Delta\hat{\sigma}^{ab\to cdX^\prime}$ the relevant, perturbatively calculable 
hard partonic cross sections at next-to-leading order (NLO) accuracy. 
The subscripts $(++)$ and $(+-)$ denote the helicities
of the lepton beam and the nucleon target at rest. 
$S$ is the total center-of-mass system (c.m.s.) energy, i.e.~$S=(p_l+p_N)^2$.
The $\Delta f^N(x_b,\mu_f)$ are the usual 
spin-dependent parton densities for parton $b$ in a nucleon at a momentum fraction $x_b$ 
and scale $\mu_f$. 
$D_{c/d}^{H_{c/d}}(z_{c/d},\mu_f^\prime)$ describe the fragmentation of a parton $c/d$ into 
a hadron $H_{c/d}$ at a momentum fraction $z_{c/d}$ and scale $\mu_f^\prime$. 
$\Delta f^l(x_a,\mu_f)$ represents the spin-dependent Weizsäcker-Williams 
equivalent 
photon spectrum \cite{wwspectrum} describing the collinear radiation 
of a photon with momentum fraction
$x_a$ and virtuality lower than some upper value $Q^2_\mathrm{max}$. All 
phenomenological 
studies have been done for the so-called \textit{direct} case, where the photon 
interacts directly with a parton of the nucleon. No \textit{resolved} photon 
contributions are included so far. 
For a proper treatment of the collinear, infrared and ultraviolet 
divergencies appearing in NLO calculations 
of the hard partonic cross sections we introduced a 
variable $z$ defined by
\begin{equation}
z\equiv -\frac{\vec{p}_{T,c}\cdot \vec{p}_{T,d}}{p_{T,c}^2}
\label{definition}
\end{equation}
in a system in which the incoming beam defines the longitudinal axis. 
For a covariant definition of the variable $z$ and some technical details we 
refer 
to a work by Aurenche et al. \cite{aurenche1}. 
To guarantee that the two hadrons are in opposite hemispheres, we restrict ourselves 
to the range $z>0$. 
Needless to say, the required spin-averaged cross section $d\sigma$ 
is straightforwardly 
obtained by replacing all polarized quantities in Eq. (\ref{xsec}) by their appropriate unpolarized
counterparts.

\section{Phenomenological results}

In our phenomenological studies we concentrate on the production of charged 
hadrons
made of light quark flavors. We sum over pions, kaons, and (anti-)protons and use 
fragmentation functions of KKP \cite{kkp}. For parton distributions we 
employ the unpolarized CTEQ6M \cite{cteq} and polarized 
GRSV standard sets \cite{grsv} as well as the sets of DNS \cite{dns}. 
If it is not stated otherwise, 
the factorization/renormalization scales in Eq. (\ref{xsec}) are all set equal $\mu\equiv\mu_f=\mu_f^\prime=\mu_r=p_{T,c}+p_{T,d}$.
All NLO results presented in this paper are preliminary.
\subsection{Two-hadron production at COMPASS}
At the COMPASS experiment at CERN polarized muons are scattered with a beam 
energy of 
$E_\mu=160\;\mathrm{GeV}$ off the deuterons in a polarized $^6\mathrm{LiD}$ 
solid-state target corresponding to a c.m.s.~energy of 
$\sqrt{S}\simeq 18\;\mathrm{GeV}$. 
For the calculations we demand that hadron $H_c$ has a
\begin{wrapfigure}{lh}{0.49\columnwidth}
\includegraphics[width=0.5\columnwidth]{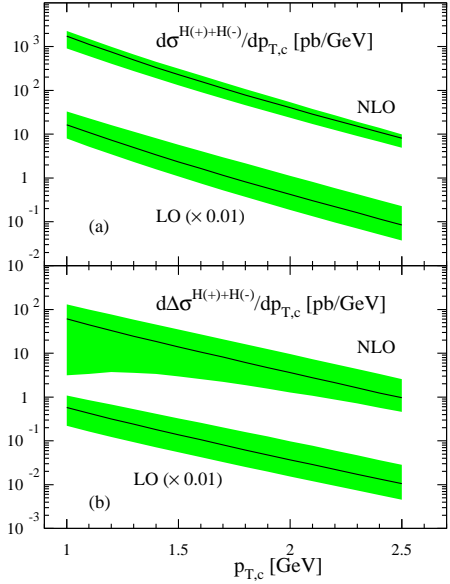}
\caption{Scale dependence at COMPASS kinematics.}
\label{Fig:scadepcomp}
\end{wrapfigure}  
scattering angle less than $\theta_\mathrm{max}=70\;\mathrm{mrad}$, 
since it was pointed out in an earlier work \cite{pairs} that resolved photon 
contributions 
become more dominant for the full acceptance of COMPASS,
$\theta_\mathrm{max}=180\;\mathrm{mrad}$.
No acceptance cut can be implemented for the other hadron $H_d$ due 
to the variable $z$ defined in Eq. (\ref{definition}). The fraction $x_a$ 
of the lepton's momentum 
taken by the photon is restricted to be in the range $0.1<x_a<0.9$, whereas the maximal 
virtuality $Q_\mathrm{max}^2$ in the Weizsäcker-Williams spectrum is 
$Q_\mathrm{max}^2=0.5\;\mathrm{GeV}^2$. The fractions of the parton's momenta 
carried by the produced hadrons are restricted to the range $z_c,z_d\geq0.1$.

Figure \ref{Fig:scadepcomp} shows the scale dependence of the unpolarized (a)
and polarized (b) cross section for 
COMPASS kinematics at LO and NLO accuracy with a cut $z>0.6$ on the partonic level. 
For the polarized parton distributions we employed DNS Set1.
The shaded 
bands indicate the resulting scale uncertainty of the cross sections when 
varying the scales in the range $1/2(p_{T,c}+p_{T,d})\leq \mu\leq 2(p_{T,c}+p_{T,d})$. 
At NLO the scale dependence is reduced in the unpolarized cross section. 
This might indicate that we are in a regime where perturbative QCD 
is applicable. However {\em{no}} improvement is observed in the polarized 
case and in both the unpolarized and polarized case for lower z-cuts 
($z>0.2,0.4$). As a consequence the  applicability of the perturbative approach
can not be taken for granted. 
An important benchmark for testing the pQCD framework would be, for instance, 
to check if the unpolarized data fall within the uncertainty band, 
where the parton distributions are already known very well. 

Once the applicability of pQCD is established, the double spin asymmetry turns out to be 
very sensitive to the gluon polarization assumed in the calculation. 
Varying from maximal positive to maximal negative sets of 
GRSV the asymmetry is in the range $-0.1\lesssim A_{LL}\lesssim0.4$ which would allow at
least a determination of the sign of $\Delta g$.
\subsection{Two-hadron production at HERMES}
At the HERMES experiment at DESY a longitudinally polarized  electron  
(positron) beam with $E_e\simeq 27.5\;\mathrm{GeV}$ is scattered off a proton or 
deuterium gas target. We concentrate on results for a deuterium target 
as this has the better statistics. The corresponding c.m.s.~energy 
is $\sqrt{S}\simeq7.25\;\mathrm{GeV}$. We choose a maximal photon virtuality of 
$Q^2_\mathrm{max}=0.1\;\mathrm{GeV}^2$ and restrict the momentum fraction
$x_a$ of the lepton carried by the produced photon to $0.2\leq x_a\leq 0.9$.
For hadron $H_c$ we use an acceptance cut of
$40\;\mathrm{mrad}<\theta_\mathrm{lab}<220\;\mathrm{mrad}$.
The fraction of the parton's momenta carried by the produced hadrons 
are restricted to the range $z_c,z_d\geq0.1$.
\begin{wrapfigure}{t}{0.49\columnwidth}
\includegraphics[width=0.48\columnwidth]{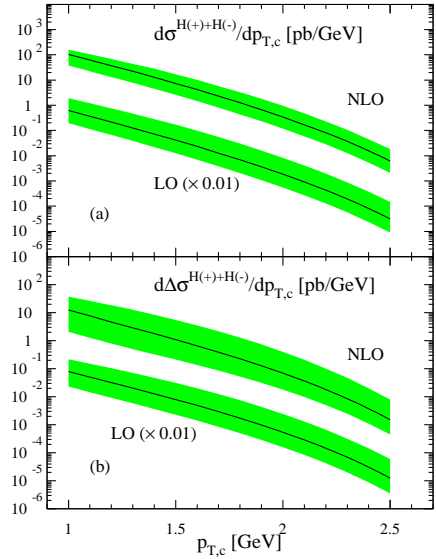}
\caption{Scale dependence at HERMES kinematics.}\label{Fig:scadepherm}
\end{wrapfigure}
Figure \ref{Fig:scadepherm} shows the scale dependence of the unpolarized (a) 
and polarized (b) cross 
sections as a function of the transverse momentum of one hadron $p_{T,c}$ when 
varying the scales in the range $1/2(p_{T,c}+p_{T,d})\leq \mu\leq 2(p_{T,c}+p_{T,d})$.
Due to the smaller c.m.s. energy there is no noticeable reduction of the 
scale dependence when 
going to NLO accuracy. We observed the same for other $z$-cuts both for 
unpolarized and polarized case. We emphasize that all comments about 
applicability 
of pQCD also apply here.
Again, as it is for COMPASS kinematics, the double spin asymmetry $A_{LL}$ 
is very sensitive to the chosen $\Delta g$ polarization: 
$-0.1\lesssim A_{LL}\lesssim 0.5$ when varying between $\Delta g=g$ and 
$\Delta g=-g$ at the input scale 
with $g$ the unpolarized gluon distribution in the nucleon.

\section*{Acknowledgments}
C.H. was supported by a grant of the ``Bayerische Eliteförderung''. 
This work was supported in part by the ``Deutsche Forschungsgesellschaft 
(DFG)''.


\begin{footnotesize}

\end{footnotesize}


\end{document}